# Effects of laser polarization in laser-assisted electron-helium inelastic collisions: a sturmian approach

**O. El Akramine, A. Makhoute and D. Khalil**

U.F.R de Physique Atomique, Molèculaire & Optique Appliquée,

Faculté des Sciences, Université Moulay Ismaïl,

B.P. 4010 Beni M'hamed, Meknès, Morocco.

**A. Maquet and R. Taïeb**

Laboratoire de Chimie Physique-Matière et Rayonnement,

Université Pierre et Marie Curie, 11 Rue Pierre et Marie Curie,

75 231 Paris Cedex 05, France.

**Abstract**

The influence of linearly and circularly polarized laser fields on the dynamics of fast electron-impact excitation in atomic helium is discussed. A detailed analysis is made in the excitation of $2^1S$, $3^1S$ and $3^1D$ dressed states of helium target. By using a semiperturbative treatment with the Sturmian basis expansion, we take into account the target atom distortion induced by a laser field. Important differences appear between the angular distributions depending on the different states of polarization, in particular the circular polarization presents an experimental interest. We give new features (intermediate resonances) for both polarizations, concerning the n = 2 states of helium for emission and the n = 3 for the absorption, in term of laser frequency. Qualitative differences from the case of laser-assisted elastic collisions have been evidenced.

# 1. Introduction

In the recent years, the study of electron–atom collisions in the presence of a laser field is a subject of intense research activity, not only because the importance of these processes in applied domains (such as plasma heating), but also in view of their interest in fundamental atomic collision theory. Experimentally, laser-assisted electron-atom scattering processes have recently become feasible. Several experiments have been performed, in which the exchange of one or more photons between the electron–target and the laser field has been observed in laser–assisted elastic [1] and inelastic scattering [2-5]. In particular, the excitation processes have been largely investigated in the literature by several authors [6-9], mainly in the perturbative (weak-field) limit. The first theoretical studies on the inelastic scattering were inspired from the pioneering works [10-12], in which the interaction between the free electron and the field can be treated exactly (i.e. to all orders in the field strength) by using the exact Volkov waves [13].

The incorporation of laser parameters as intensity, frequency and in particular polarization in the laser–assisted collisions gives interesting results and considerably enrich the study of the collision process. The influence of this later parameter have attracted a great deal of attention in theoretical works and experimental. In fact the theoretical studies of polarization dependence have been previously performed by Cavaliere et al. [14] for Simultaneous electron–photon excitation at high impact energies and large differences have been predicted. For impact energies near threshold there is experimental evidence of differences between linear and circular polarization [15] and a recent extension of the Kroll and Watson theory by Mittleman [16] predicts differences in the first order of the development of the transition amplitude as a function of the laser frequency. Taïeb et al [17] studied the influence of the laser polarization on the angular distribution of the ejected electron in laser-assisted (e, 2e) reactions. Fainstein and Maquet [18] studied the polarization dependence of laser–assisted electron–hydrogen elastic collisions, where important differences appear between the angular distributions depending on the different states of polarization.

In the present paper we want to extend our previous results [22] to the case of laser-assisted inelastic electron-helium collisions. A comparison between the two polarizations linear and circular of electric laser field will be made for different geometries. We have performed an ''exact'' evaluation of the needed infinite sum-over-states, based on simplified hydrogenic functions of the excited spectrum of helium [12]. In the purpose to confirm our numerical results, we have performed the calculations as in our previous paper [22] by two different methods both based on the Sturmian basis expansion. The first one consists in expanding the radial term of the Coulomb Green function on a discrete basis of Sturmian functions [19-20], which allows us to take into account exactly the bound-continuum-state contributions, which is of crucial importance for electron impact excitation at intermediate energies [23]. In the second method, the calculation is performed by expanding the first-order perturbed wave function onto the same Sturmian basis [17, 21]. The use of these two methods independently allows us to accurately determine the contribution of the entire singly discrete or continuous excited spectrum (note that the doubly excited states are not taken into account by these methods). The present technique has been applied extensively to a variety of laser assisted electron atom collisions involving the transfer of one or several photons between the electron-atom system and the laser field in the cases of elastic, inelastic collisions and (e,2e) reactions. We neglect the exchanges effects in the presence work, since the field-free exchange effects are essentially negligible at the high impact energies considered here, and they are either smaller or slightly enhanced in the presence of a laser field [24].

The paper is structured as follows. In section 2 we present the general formalism of laser-assisted inelastic electron-helium collisions in the case of linear and circular polarizations. An account is then given of the techniques that we have used to evaluate the scattering amplitudes. Section 3 contains a detailed presentation of our numerical results as well as their physical interpretations which will be discussed in this part and section 4 concludes the paper. Atomic units used through this paper.

## 2- Theory

Following our previous work [22], we consider the classical monochromatic and single mode field, that is spatially homogenous over atomic dimensions and has, in the Coulomb gauge, the electric field represented in the collision plane $(\hat{x} - \hat{y})$

$$\mathbf{F}(t) = F_0 \left[ \hat{x} \sin(\omega t) - \hat{y} \cos(\omega t) \tan(\eta/2) \right] \quad (1)$$

where $F_0$ and $\omega$ are respectively the amplitude and the frequency of the electric laser field. The $\eta$ parameter measures the ellipticity of the field polarization and we have the particular cases of linear polarization ($\eta = 0$) and circular polarization ($\eta = \pi/2$). We can represent the electric laser field in terms of their spherical components by

$$\mathbf{F}(t) = F_0 \sum_{\lambda = \pm 1} i\,\lambda\,\hat{\varepsilon}_\lambda \exp(-i\lambda(\omega t + \varphi)) \quad (2)$$

where $\hat{\varepsilon}_\lambda = \hat{x} + \lambda\,i\,\hat{y}\,\tan\left(\eta/2\right)$ is the unitary polarization vector.

In the presence of this field, we consider the inelastic scattering process (electron–helium), represented by the following equation

$$e^-(\mathbf{k}_i, E_{k_i}) + He(1^1S) + L\omega \rightarrow He^* + e^-(\mathbf{k}_f, E_{k_f}) \quad (3)$$

where $\mathbf{k}_i$ and $\mathbf{k}_f$ are respectively the wave vectors of the incident and scattered electrons in the presence of the laser field. $E_{k_i} = \frac{k_i^2}{2}$ and $E_{k_f} = \frac{k_f^2}{2}$ are the projectile initial and final energies. The target helium is initially in the ground state $1^1S$ and will be excited after the scattering in one of the ''bound'' final states. The integer L is the number of photons transferred between the (electron–helium) system and the field, positive values of L correspond to the absorption of photons by the system and negative ones correspond to stimulated emission of photons.

The energy conservation relation corresponding to the laser–assisted excitation (Eq. 3), reads

$$E_{k_i} + E_0^{He} + L\omega = E^* + E_{k_f} \tag{4}$$

where $E_0^{He}$ = -2.904 a.u. and $E^*$ are, respectively, the ground and the final excited state energies of the helium target.

The interaction between the projectile and the laser field is treated exactly and his solution is given by the non–relativistic Volkov wave function $\chi_k(\mathbf{r}_0, t)$ [13, 22], where $\mathbf{k}$ being the projectile wave vector and, where $\mathbf{r}_0$ represents free electron coordinate.

For the interaction laser–target, we are interested by fields which have electric strengths smaller than the atomic unit ($F_0 \ll 5.10^9$ V/cm), and frequencies different from the atomic transition energies, then the perturbation theory is the most appropriate method to solve the interaction process. At first order time–dependent perturbation theory the '' dressed '' wave function $\Phi_n(\mathbf{X}, t)$ is well known (see ref. [22]), where $\mathbf{X} \equiv (\mathbf{r}_1, \mathbf{r}_2)$ are the coordinates of the two helium target electrons and n is the principal quantum number.

Remembering that we are working in the geometry, where the incident electron is fast and exchange effects are small, we shall, as a first approximation, carry out a first–Born treatment of the scattering process. The first–Born S–matrix element for the inelastic collision from the ground state of the target to a final excited state of energy $E^*$, in the direct channel, is given by

$$S_{f,0}^{B1} = -i \int_{-\infty}^{+\infty} dt \left\langle \chi_{k_f}(\mathbf{r}_0, t) \Phi_f(\mathbf{X}, t) \left| V_d(\mathbf{r}_0, t) \right| \chi_{k_i}(\mathbf{r}_0, t) \Phi_0(\mathbf{X}, t) \right\rangle \tag{5}$$

where $V_d(\mathbf{r}_0, t) = -\dfrac{2}{r_0} + \sum_{j=1}^{2} \dfrac{1}{r_{0j}}$, is the direct interaction potential, $r_{0j} = |\mathbf{r}_0 - \mathbf{r}_j|$, $\chi_{K_i}(\mathbf{r}_0, t)$ and $\chi_{K_f}(\mathbf{r}_0, t)$ are respectively the Volkov wave functions of the incident and scattered electrons in the presence of the laser field. $\Phi_0(\mathbf{X}, t)$ and $\Phi_f(\mathbf{X}, t)$ are respectively the " dressed " atomic wave functions describing the fundamental and final excited states. After integration on the time variable, we have

$$S_{f,0}^{B1} = (2\pi)^{-1} i \sum_{L=-\infty}^{+\infty} \delta(E_{k_f} - E_{k_i} + E^* - E_0^{He} - L\omega) e^{iL\gamma_k} f_{f,0}^{B1,L}(\mathbf{K}) \quad (6)$$

with

$$\tan(\gamma_k) = \frac{\mathbf{k}.\hat{\mathbf{y}}}{\mathbf{k}.\hat{\mathbf{x}}} \tan\left(\eta/2\right) \quad (7)$$

$\mathbf{K} = \mathbf{k}_i - \mathbf{k}_f$ is the momentum transfer which is relatively small, $f_{f,0}^{B1,L}$ is the first–Born approximation to the inelastic scattering amplitude with the transfer of L photons, which can be written as

$$f_{f,0}^{B1,L}(\mathbf{K}) = f_{elec}^{B1,L}(\mathbf{K}) + f_{atom}^{B1,L}(\mathbf{K}) \quad (8)$$

with

$$f_{elec}^{B1,L}(\mathbf{K}) = -\frac{2}{\mathbf{K}^2} J_L(R_K) \langle \psi_f | \tilde{V}_d(\mathbf{K},\mathbf{X}) | \psi_0 \rangle \quad (9)$$

and

$$f_{atom}^{B1,L}(\mathbf{K}) = f_{II} + f_{III} \quad (10)$$

where

$$f_{II} = \left(-\frac{2}{K^2}\right) \frac{i}{2} \sum_n \left[ \frac{J_{L+1}(R_K) e^{i\gamma_k} M_{n0}^-}{\omega_{n0} + \omega} - \frac{J_{L-1}(R_K) e^{-i\gamma_k} M_{n0}^+}{\omega_{n0} - \omega} \right] \langle \psi_f | \tilde{V}_d(\mathbf{K},\mathbf{X}) | \psi_n \rangle \quad (11)$$

and

$$f_{III} = \left(-\frac{2}{K^2}\right)\left(-\frac{i}{2}\right) \sum_n \left[ \frac{J_{L-1}(R_K) e^{-i\gamma_k} M_{fn}^+}{\omega_{nf} + \omega} - \frac{J_{L+1}(R_K) e^{+i\gamma_k} M_{fn}^-}{\omega_{nf} - \omega} \right] \langle \psi_n | \tilde{V}_d(\mathbf{K},\mathbf{X}) | \psi_0 \rangle \quad (12)$$

The terms $f_{elec}^{B1,L}(\mathbf{K})$ and $f_{atom}^{B1,L}(\mathbf{K})$ are called, respectively ''electronic'' (corresponds to the interaction of the laser field with incident electron only) and ''atomic'' (include the dressing effects and thus describe the distortion of the atom target by the electromagnetic radiation), with $M_{n'n}^{\pm} = F_0 \langle \psi_{n'} | \hat{\boldsymbol{\varepsilon}}_{\pm}.\mathbf{R} | \psi_n \rangle$ are the dipole matrix elements, $\omega_{nn'} = E_n - E_{n'}$ are the atomic transition frequencies and the potential $\tilde{V}_d(\mathbf{K}, \mathbf{X})$ is given by

$$\tilde{V}_d(\mathbf{K}, \mathbf{X}) = \sum_{j=1}^{2} \exp(i\mathbf{K} \cdot \mathbf{r}_j) - 2 \qquad (13)$$

In the equations (11) and (12) $\psi_n$ is a target state of energy $\omega_n$ in the absence of external field and

$$R_k = \alpha_0 \left[ (\mathbf{k} \cdot \hat{\mathbf{x}})^2 + (\mathbf{k} \cdot \hat{\mathbf{y}})^2 \tan^2\left(\eta/2\right) \right]^2 \qquad (14)$$

where $\alpha_0 = \dfrac{F_0}{\omega^2}$ represents the oscillation amplitude of a classical electron in a laser field, $J_L$ is an ordinary Bessel function of order L and $\mathbf{R} = \sum_{j=1}^{2} \mathbf{r}_j$ is the sum of all target coordinates.

The first–Born differential cross section for the helium excitation with the transfer of L photons is given by

$$\left(\frac{d\sigma_{inel}}{d\Omega}\right) = \frac{k_f}{k_i} \left| f_{f,0}^{B1,L}(\mathbf{K}) \right|^2 \qquad (15)$$

The corresponding first–Born differential cross section for the helium excitation with the dressing neglected reads

$$\left(\frac{d\sigma_{inel}}{d\Omega}\right)_{no\,dressing} = \frac{k_f}{k_i} \left| f_{elec}^{B1,L}(\mathbf{K}) \right|^2 \qquad (16)$$

In the calculation of the two amplitudes of the equation (10), we need to know the explicit form of the atomic wave functions. For the ground state of helium, we use the wave function proposed by Byron and Joachain [25]. For the $2^1S$, $3^1S$ states, Francken et al. [12] have constructed the corresponding wave functions. The form of the states $n^1P$ and $n^1D$ is also given by Francken et al. [12].

In the case of the excitation of the $2^1S$, $3^1S$ and $3^1D$ states, we have only the $n^1P$ as intermediate states; it is a simple matter to include all simply excited states of this type. In the case of the excitation of the $n^1P$ states, we have taken into account besides the $n^1D$ states, the $n^1S$ with principal quantum number $n \leq 3$ only. We note that the doubly excited target states are neglected because the weak contribution of these states for the inelastic processes [12].

The second-order hydrogenic matrix elements appearing in the equations (11) and (12) have the general form

$$T_{1,2}^{\pm} = \langle \phi_f(\mathbf{r}) | V_1(r) G_c(\Omega_{1,2}^{\pm}) V_2(r) | \phi_0(\mathbf{r}) \rangle \qquad (17$$

where $V_1$ and $V_2$ are any perturbation operators, which takes the following forms, namely $e^{i\mathbf{K}\cdot\mathbf{r}}$ and $\hat{\boldsymbol{\varepsilon}}_{\pm}\cdot\mathbf{r}$.

with

$$\Omega_1^{\pm} = E^* - E_0^{He^+} \pm \omega \qquad (18)$$

and

$$\Omega_2^{\pm} = E^{He} - E_0^{He^+} \pm \omega \qquad (19)$$

where $E_0^{He^+} = -2.\,\mathrm{a.u.}$ $\phi_f(\mathbf{r})$ and $\phi_0(\mathbf{r})$ are respectively the orbital functions of the final and the initial states. The main difficulty encountered in the numerical estimation of the transition amplitude lies in the computation of the second order atomic matrix elements $T_{1,2}^{\pm}$ containing the Coulomb Green function $G_c(\Omega_{1,2}^{\pm})$, especially when the argument $\Omega_{1,2}^{\pm}$ of the Coulomb Green's function is positive. These difficulties can often be overcome by using an extension of Zernik's approach [26] for the solution of the relevant inhomogeneous differential equations or by using partial–wave expansions of the amplitudes and the corresponding radial parts of the partial–wave component $T_{1,2}^{\pm}$ which are given respectively by the following expression

$$M_{1,2}^{\pm} = \left\langle R_f^{He}(r) \middle| V_2^r(r) G_\lambda(\Omega_{1,2}^{\pm}) V_1^r(r) \middle| R_0^{He}(r) \right\rangle \qquad (20)$$

Therefore the calculation of the radial amplitudes (17) reduces to the computation of matrix elements of the general form (Eq. 20). Where $V_1^r$ and $V_2^r$ are any radial perturbation operators corresponding to $j_\lambda(Kr)$ and r. With $j_\lambda(Kr)$ is a spherical Bessel function, $G_\lambda(\Omega_{1,2}^{\pm})$ is the radial part of the λth partial–wave component of the Coulomb green function, and $R_f^{He}(r)$ and $R_0^{He}(r)$ are respectively the radial functions of the final and initial states.

The radial amplitude (20) can be conveniently calculated by using a Sturmian approach. In order to double–check our numerical results we have used two different methods, which although relying both a Sturmian approach, differ somewhat in the practicalities of the computation. These methods are namely:

i) Sturmian expansion of the Coulomb Green function [19-20].

ii) Sturmian expansion of the first–order perturbed wave function [17, 21, 22].

The basic idea underlying of the two techniques that we have used lies in the expansion onto the radial Sturmian basis of one or several components of the general amplitude $M_{1,2}^{\pm}$.

## 3. Results and discussion

The present semiperturbative method with the Sturmian basis expansion takes into account the target atom distortion induced by the presence of laser field. The validity of our treatment is based on the fact that the laser–helium target interaction is nonresonant. We note that the excitation process can be considered as nonresonant if for a given frequency, the intensity does not exceed a certain limit [12]. The condition on the intensity is more stringent if the laser frequency is comparable to any characteristic atom excitation frequency. Such conditions will be respected by our choice of the Nd–YAG laser of frequency $\omega = 1.17$ eV and electric intensity $F_0 = 10^6$ V/cm.

We are interested in demonstrating the effects of the polarization effects in the inelastic collision of fast electrons by a helium target in the presence of a laser field. For linear polarization, we have considered two particular geometries where the polarization vector of the field is taken to be parallel to the momentum transfer **K** ($\mathbf{F}_0 // \mathbf{K}$), and to be parallel to the wave vector of the incident electron $\mathbf{k}_i$ ($\mathbf{F}_0 // \mathbf{k}_i$). For circular polarization, we have chosen two distinct geometries corresponding respectively to the laser wave vector **k** being perpendicular to the scattering plane (CPP) and to be parallel to the scattering plane (CPC). Note that, for linear polarization the laser–assisted differential cross section only depends on the orientation of the polarization unit vector $\hat{\boldsymbol{\varepsilon}}_\pm$.

In order to illustrate the effects of the laser polarization on the variations of the differential cross sections, we have chosen to compare our results obtained for $\mathbf{F}_0 // \mathbf{K}$ (linear polarization figure 1-a) with those corresponding to CPP (circular polarization figure 1-c). The reason for this particular choice is that the electronic term $f_{elec}^{B1,L}(\mathbf{K})$, is the same for both geometries, because the argument of the Bessel function reduces to an identical value $R_k = \alpha_0 K$ in these two cases. The same situation occurs when one compares the differential cross sections corresponding to $\mathbf{F}_0 // \mathbf{k}_i$ ( linear polarization, figure 1-b) with those obtained for CPC (circular polarization coplanar, figure 1-d), the argument of the Bessel function being then reduced to an identical value $R_k = \alpha_0 (k_i - k_f \cos(\theta))$, where $\theta$ is the scattering angle. In both cases, the differences observed in the angular dependence of the cross sections result from the differences between the contributions of the atomic terms, i.e. on the dressing of the target.

Before presenting the results of our calculations, we want to make a remark concerning the phase $\gamma_k$ used in equations (6), (7), (11) and (12) for taking into account the effects of the laser polarization on the variations of the laser-assisted differential cross sections. This phase is particularly important for circular polarization CPC, when $\tan(\gamma_k)$ vanishes, so that $\gamma_k = 0 \,[\mod \pi]$, these values of $\gamma_k$ correspond to the case where the two components of the electric field in the plane $(\hat{\mathbf{y}}, \hat{\mathbf{z}})$ vary, as a function of time, with the same phase $\gamma_k = 0$, or with opposite phases $\gamma_k = \pi$. The change of phase is absent in the case of circular polarization CPP.

The figures 2 and 3 represent the first-Born differential cross sections corresponding to the excitation of the $2^1S$ and $3^1D$ states as function of scattering angle $\theta$, for the absorption of one photon $L = 1$ (inverse bremsstrahlung process), the fast incident electron is characterized by an energy of $E_{k_i} = 500$ eV. In each of these two figures, we have displayed the inelastic scattering amplitude for two different geometries for each polarization state. For linear polarization with $\mathbf{F}_0$ // $\mathbf{K}$, $\mathbf{F}_0$ // $\mathbf{k}_i$ and for circular polarization with two distinct orientations of a circularly polarized laser beam CPP and CPC.

As indicated in our previous paper for elastic scattering [22], we have observed the existence of two kinds of minima ($m_1$) and ($m_2$) corresponding, respectively, to the situations when $f_{elec}^{B1,L}(\mathbf{K}) + f_{atom}^{B1,L}(\mathbf{K}) = 0$, and at angles such that the argument $R_k$ of the Bessel functions actually vanishes. This last minimum exists only in the case when $\mathbf{F}_0$ // $\mathbf{k}_i$, it localisation in terms of $\theta$ is given by the relation $k_i - k_f \cos(\theta) = 0$ and can be observed, in the case of excitation, for large laser frequencies, which are not feasible in practice. So all destructive interferences, presented in this paper, correspond to the minimum ($m_1$).

In figure 2(a) and figure 2(b), we show the laser-assisted differential cross sections corresponding to $1^1S \rightarrow 2^1S$ excitation process. The complete result obtained by using the scattering amplitude Eq.(10) for two polarizations is compared to the ''electronic'' cross section in which dressing effects are neglected. As in the case of elastic collisions [22] dressing effects are shown to be dominant in the forward direction for linear polarization where $\mathbf{F}_0$ // $\mathbf{K}$ and $\mathbf{F}_0$ // $\mathbf{k}_i$, and circular polarization (CPC) with $\gamma_k = \pi$ and for larger scattering angles for circular polarization CPP and CPC with $\gamma_k = 0$. We can see that for angles below 12° there are important differences between the two polarizations. This behaviour is particularly important from the experimental point of view since it is in principle easier to measure the laser–assisted differential cross section amplitudes (for larger scattering angles), where the dressing effects of the target contribute significantly. Moreover, we notice a destructive interference between the electronic and the atomic amplitudes near of $\theta \approx 3, 2°$, for linear polarization $\mathbf{F}_0 // \mathbf{K}$ and $\mathbf{F}_0 // \mathbf{k}_i$ and near of $\theta \approx 7°$ for the circular polarization CPC with $\gamma_k = \pi$, (the electronic and the atomic amplitudes are varying in opposite directions when the momentum transfer increases). The presence of such interference pattern is a general feature of $1^1S \rightarrow n^1S$ transitions in the case of inverse bremsstrahlung $L > 0$. This is due to the presence, in the atomic term, of S-P transition amplitudes, which behave like $\mathbf{K}^{-1}$ for small $\mathbf{K}$.

The results displayed in figures 3(a) and 3(b) correspond to the $1^1S \rightarrow 3^1D$ transition and show that dressing effects are also dominant for scattering angles $\theta < 15°$, for both linear and circular polarizations. Moreover, we note that in this case the interference between the electronic and atomic amplitudes is constructive. This contrasts with the results obtained for $1^1S \longrightarrow n^1S$ transitions. We notice that in the case of $3^1S$ state excitation, a qualitatively similar behaviour, as in the case of $1^1S \longrightarrow 2^1S$ process, is observed. We do not show figures concerning $1^1S \rightarrow n^1P$ transitions, since dressing effects are rather small in this case. Indeed the electronic S-P amplitude, which behaves like $\mathbf{K}^{-1}$ for small $\mathbf{K}$, now dominates the cross section at small angles.

In the case of emission process $L = -1$, figure 4 shows an opposite behaviour of the cross section than for the case of absorption $L = 1$, in fact, the destructive interference observed in figure 2(a) and 2(b), for the two geometries $\mathbf{F}_0 // \mathbf{K}$ and $\mathbf{F}_0 // \mathbf{k}_i$ of the linear polarization, becomes constructive, and the constructive interference of the cross section for circular polarization becomes destructive. This behaviour can be explained by the change of Bessel functions from absorption $L = 1$ to emission $L = -1$, making a change of sign of the atomic amplitude.

In figures 5, 6 and 7, we present the frequency dependence of the differential cross sections for the cases of inverse bremsstrahlung ( absorption of one photon $L = 1$) and stimulated bremsstrahlung (emission of one photon $L = -1$). We take a fixed scattering angle $\theta = 10°$ and an incident electron energy $E_{k_i} = 500$ eV. We have normalized the cross sections, as function of laser frequency, to the averaged laser intensity $I = \dfrac{cF_0^2}{8\pi}$.

Figure 5(a) displays the excitation of $1^1S \rightarrow 3^1S$ for absorption of one photon and for the two polarizations (linear polarization with $\mathbf{F}_0 // \mathbf{K}$ and circular polarization CCP). At a frequency of $\omega \approx 13$ eV, we remark a minimum for the circular polarization CPP, which is induced by destructive interferences between electronic and atomic terms of the differential cross section. Indeed, the phase-dependent factor $e^{\pm i\gamma_k}$ present in the atomic term changes the sign of its real part thus making the change from constructive to destructive interference. This type of minimum can not exist in the linear polarization where $\mathbf{F}_0 // \mathbf{K}$ because, here we have a constructive interference. For the large frequencies, abrupt changes in the vicinity of Bohr transition frequencies indicate that the behaviour of the cross sections with respect to the laser frequency strongly depends on the structure of the target. It is interesting to remark the presence of minima between two successive resonances for linear and circular polarization. This behaviour results from the fact that the resonant atomic amplitudes change of sign between two resonances and can compensate the direct contribution independently of the polarization state. It is also important to note that these scattering amplitudes are sensitive to the presence of those Bohr transitions even far away from resonance. Figure 5(b) displays the differential cross sections as a function of the laser frequency, for linear polarization where $\mathbf{F}_0 // \mathbf{k}_i$ and CPC circular polarization where $\gamma_k = 0$, we remark a qualitatively similar behaviour as in figure 5(a), with a small shift of the minimum for the CPC circular polarization with $\gamma_k = 0$. While in the case where $\gamma_k = \pi$, the change of sign of atomic amplitude, induced by the phase factor $e^{\pm i\gamma_k}$ introduces a constructive interference. We note that for the CPC polarization with $\gamma_k = \pi$, the cross section dependence in terms of frequency shows a similar behaviour as in terms of scattering angle. An important point in this figure 5 is the observation of intermediate resonances, which correspond to transitions involving intermediate states. Such peaks can be interpreted as corresponding to Simultaneous Electron -Photon Excitation (SEPE), (see Appendix). This process (SEPE), which is investigated in the excitation of the helium $2^3S$ state [27, 28], can be explained by an excitation which occur through the 'simultaneous' impact of an electron and one or more photons. Indeed, this excitation is accomplished by the absorption or the emission of one photon energy $\hbar\omega$ combined with a simultaneous inelastic scattering in which the electron provides the energy decrement $(E_{k_f} - E_{k_i})$ required to excite the desired state. We note that these peaks appear in the region of small frequencies.

The results displayed in figures 6 correspond to the $1^1S \rightarrow 3^1D$ excitation as a function of laser photon energy. We observe in this case destructive interferences for the two polarizations (linear and circular) and for the different geometries. We observe also the intermediate resonances corresponding to the SEPE process.

Figure 7 represents the $1^1S \rightarrow 2^1S$ excitation cross section as a function of the laser photon energy in the case of emission process L=-1, and shows a different behaviour than for the case of absorption L = 1 (Figure 8). In fact, for the two polarizations, the scattering amplitude exhibits peaks corresponding to radiative transitions which are induced by SEPE process (see Appendix). This behaviour of the differential cross sections, at low frequencies, characterize the excitation of states with principal quantum number n=3 for absorption and the excitation of n=2 for emission in term of frequencies. Such characteristics constitute one of the main differences between elastic and inelastic scattering in a laser field. By comparing figure 7 and figure 8, we note that in the case of bremsstrahlung stimulated, a large difference is observed with respect to the case of inverse bremsstrahlung. This asymmetry does not exist in the elastic scattering case. We conclude, for these figures 5, 6 and 7 displaying the frequency dependence, that all results have common features: peaks at low frequencies corresponding to the SEPE process, minimum between 0.2 and 0.4 a.u. due to destructive interferences between electronic and atomic amplitudes and for large frequencies there are resonances induced by the presence of the Bohr transitions between the initial and intermediate states.

## Conclusion

In this paper we have extended our treatment of electron-helium elastic collisions in the presence of a linearly and circularly polarized laser field to the case of the excitation. The calculations have been performed by two different methods both based on the Sturmian basis expansion. Important differences have been found when we compare the differential cross sections for two laser polarizations (circular and linear) by using different geometries. In the cases of excitation $n^1S$ and $n^1D$, we have remarked that dressing effects are important and dominant, for linear polarization and circular polarization with $\gamma_k = \pi$, only in the region of small scattering angles, while for circular polarization with $\gamma_k = 0$, the dressing effects are important at large scattering angles. Our results show that, everything else being fixed, a circularly polarized laser (CPP and CPC with $\gamma_k = 0$) can give larger cross sections than a linearly polarized one, by several orders of magnitude. This should constitute an interesting and attractive point for the experimentalists to measure the cross section amplitudes in the case of circularly polarized laser beam. New features have been observed in the case of frequency dependence, indeed, in the case of n = 3 excitation for absorption and n = 2 for emission, intermediate resonnances appear in the region of small laser frequencies, which correspond to dipole transitions interpreted by Simultaneous Electron -Photon Excitation process.

## Acknowledgements

We wish to thank Professor C. J. Joachain for helpful communication. It is a pleasure to thank also Professor B. Wallbank and Doctor N. J. Mason for sending us their interesting reprints concerning this subject. The Co-operation between ' Université Moulay Ismaïl, Meknès, Maroc ' and ' Université Pierre et Marie Curie, Paris VI, France ' has greatly influenced the present paper.

# APPENDIX: ENERGY DIAGRAMS CORRESPONDING TO THE SIMULTANEOUS ELECTRON-PHOTON EXCITATION OF HELIUM $2^1S$ and $3^1S$ STATES.

We interpret the resonances appearing in the region of small frequencies by representing the energy diagrams corresponding to these 'simultaneous' electron-photon excitation of n=2 for emission and n = 3 for absorption in terms of frequencies. We consider the responsible quantities of these resonances, which are given (from Eq. 16) by

$$\omega_{nf} \pm \omega = 0 \qquad (1)$$

In the other hand, the energy conservation equation (Eq. 5) reads

$$E_{k_i} + E_0^{He} + L\omega = E^* + E_{k_f} \qquad (2)$$

### 1. Case L = -1 (emission).

In this case, only the helium (n = 2) states can present such resonances. We consider the excitation of $2^1S$, and for the first resonance corresponding to the intermediate state $2^1P$, the relations (1) and (2) become

$$\omega_{2^1S} = \omega_{2^1P} - \omega$$

and

$$E_{2^1P} = E_{ki} - E_{kf} + E_{1^1S}$$

the corresponding energy diagram is

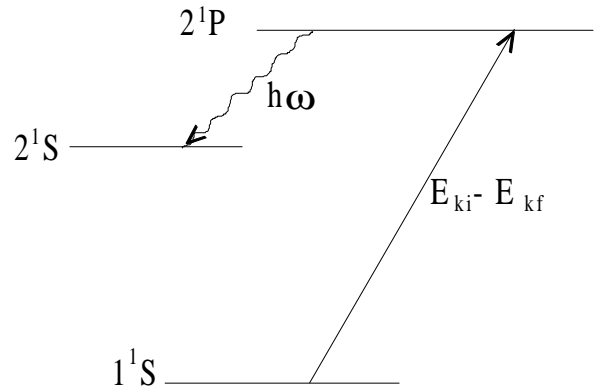

Energy level diagram of helium corresponding to the 'simultaneous' excitation of $2^1S$ with (L = -1). The photon energy is ($\hbar\omega$) and the excess of energy between incident and diffused electrons is ($E_{ki}$ - $E_{kf}$).

A similar procedure is made for the other resonances corresponding to the intermediate states $3^1P$, $4^1P$, $5^1P$,....

### 2. Case L = +1 (absorption).

Only the states with (n = 3), which can be excited by 'simultaneous 'electron-photon excitation (SEPE) and we choose the $3^1S$ for example, $2^1P$ is the unique intermediate state with energy lower than the final state. The equations (1) and (2) write

$$\omega_{3^1S} = \omega_{2^1P} + \omega$$

and

$$E_{2^1P} = E_{ki} - E_{kf} + E_{1^1S}$$

and the corresponding diagram is represented by

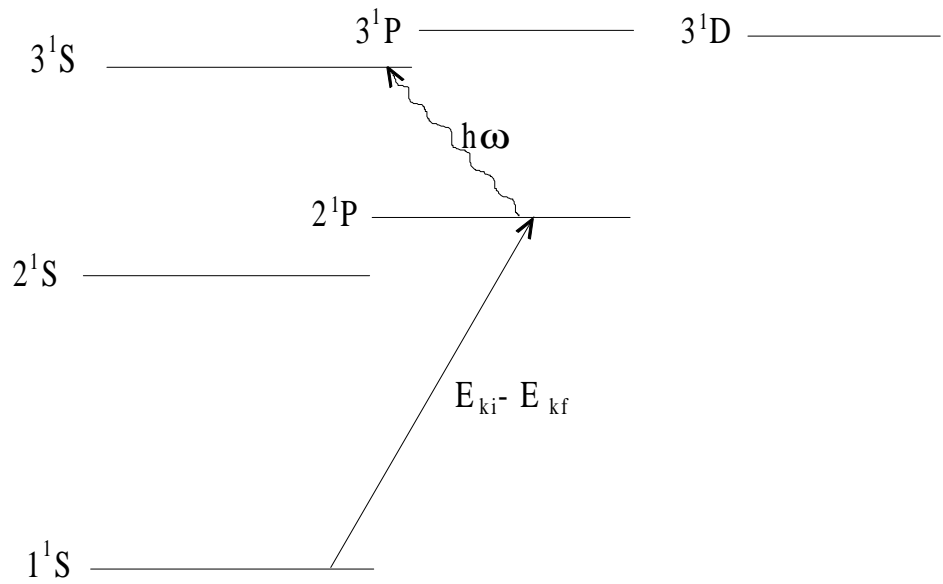

Energy level diagram of helium corresponding to the 'simultaneous' excitation of $3^1S$ with ($L = +1$). The photon energy is ($\hbar\omega$) and the excess of energy between incident and diffused electrons is ($E_{ki} - E_{kf}$).

# References

[1] A. Weingartshofer, J. K. Holmes, J. Sabbagh and S. I. Chu, J. Phys. B **16**, 1805 (1983). See also, B. Wallbank, J. K. Holmes, J. Phys. B **27**, 1221 (1994).

[2] M. A. Khaboo, D. Roundy and F. Rugamas, Phys. Rev. A **54**, 4004 (1996).

[3] S. Luan, R. Hippler and H. O. Lutz, J. Phys. B **24**, 3241(1991).

[4] N. J. Mason and W. R. Newell, J. Phys. B **22**, 777 (1989).

[5] B. Wallbank, J. K. Holmes and A. Weingartshofer, Phys. Rev. A **40**, 5461, (1989); J.Phys. B **23**, 2997 (1990).

[6] N. K. Rahman and F. H. M. Faisal, J. Phys. B **11**, 2003 (1978).

[7] S. Jetzke, F. H. M. Faisal, R. Hippler and O. H. Lutz, Z. Phys. A **315**, 271, (1984).

[8] S. Jetzke, J. Broad and A. Maquet, J. Phys. B **20**, 2887 (1987).

[9] R. S. Pundir and K. C. Mathur, Z. Phys. D **1**, 385 (1986).

[10] F. W. Byron Jr, P. Francken and C. J. Joachain, J. Phys. B **20** 5487 (1987).

[11] F. W. Byron Jr and C. J. Joachain, Phys. Rev. A **35** 1590 (1987).

[12] P. Francken, Y. Attaourti and C. J. Joachain, Phys. Rev. A **38**, 1785 (1988).

[13] D. M. Volkov, Z. Phys **94**, 250 (1935).

[14] P. Cavaliere, C. Leone and G. Ferrante; Nuevo Cimento D **4**, 79 (1984).

[15] N. J. Mason and W. R. Newell, J. Phys. B **23**, L179 (1990).

[16] M. H. Mittleman, J. Phys. B**26** 2709 (1993).

[17] R. Taïeb, V. Véniard, A. Maquet, S. Vucic and R. M. Potvielge, J. Phys. B **24**, 3229 (1991).

[18] P. D. Fainstein and A. Maquet, J. Phys. B **27**, 5563 (1994).

[19] A. Maquet, Phys. Rev. A **15**, 1088 (1977).

[20] C. J. Joachain, A. Makhoute, A. Maquet and R. Taïeb, Z. Phys. D **23**, 397 (1992).

[21] D. Khalil, A. Maquet, R. Taïeb, C. J. Joachain, and A. Makhoute, Phys. Rev.A **56**, 4918 (1998).

[22] D. Khalil, O. El Akramine, A. Makhoute, A. Maquet and R. Taïeb, J. Phys. B **31**, 1 (1998).

[23] S. Vucic, Phys. Rev. A **51**, 4754 (1995).

[24] G. Ferrante, C. Leone and F. Trombetta, J. Phys. B **15**, L475 (1982); F. Trombetta and C. J. Joachain, and G. Ferrante, ibid. **19**, 1081 (1986).

[25] F. W. Byron Jr and C. J. Joachain, Phys. Rev. A **146** 1 (1966).

[26] W Zernik, Phys. Rev. **135** A51-7 (1964).

[27] N. J. Mason and W. R. Newell, J. Phys. B **22**, 777 (1989).

[28] B. Wallbank, J. K. Holmes and A. Weingartshofer, J. Phys. B **22**, L615 (1989).

## Figures Captions

Figure 1: Selected scattering geomeries for electron-helium collisions in the presence of linearly, (a), (b) and circularly, (c), (d), polarized laser fields.
(a) $\mathbf{F}_0 // \mathbf{K}$, (b) $\mathbf{F}_0 // \mathbf{k}_i$, (c) CPP and (d) CPC see text.

Figure 2: First-Born differential cross section corresponding to the electron-impact excitation of the $2^1S$ state atomic helium with absorption of one photon (L=1) as a function of the scattering angle. The incident electron energy is 500 eV, the laser frequency is 1.17 eV and the electric field strength is $10^6$ V cm$^{-1}$. Dotted line: results obtained by neglecting the dressing of the target.
(a) Solid line: circular polarization (CPP). Dashed line: linear polarization ( $\mathbf{F}_0 // \mathbf{K}$ ).
(b) Solid line: circular polarization (CPC) with $\gamma_k = 0$ and $\gamma_k = \pi$. Dashed line: linear polarization ( $\mathbf{F}_0 // \mathbf{k}_i$ ).

Figure 3: Same as Figure 2 but for the excitation of the $3^1D$ state of atomic helium.

Figure 4: Same as Figure 2 but with emission of one photon (L=-1).

Figure 5: First-Born differential cross section corresponding to the electron-impact excitation of the $3^1S$ state atomic helium with absorption of one photon (L = 1) as a function of laser frequency for a fixed scattering angle $\theta = 10°$. The incident electron energy is 500 eV. The cross sections have been normalized to the mean laser intensity $I = \dfrac{c F_0^2}{8\pi}$. Dotted line: results obtained by neglecting the dressing of the target.
(a) Solid line: circular polarization (CPP). Dashed line: linear polarization ( $\mathbf{F}_0 // \mathbf{K}$ ).
(b) Solid line: circular polarization (CPC) with $\gamma_k = 0$. Dashed line: linear polarization ( $\mathbf{F}_0 // \mathbf{k}_i$ ).

Figure 6: Same as Figure 5 but for the excitation of the $3^1D$ state of atomic helium and with absorption of one photon (L = 1).

Figure 7: Same as Figure 5 but for the excitation of the $2^1S$ state of atomic helium and emission of one photon ( L = -1 ).

Figure 8: Same as Figure 7 but for the absorption of one photon ( L = 1 ).